\documentclass[5p, authoryear, 11pt, times]{elsarticle}

\usepackage{amssymb}
\usepackage[fleqn]{amsmath}
\usepackage{graphicx}
\usepackage{epstopdf}

\usepackage{setspace}
\usepackage{lineno}

\newcommand{\Kn}{K\!n}
\newcommand{\Ma}{M\!a}

\begin{document}

\title{Hybrid fluid/kinetic modeling of Pluto's escaping atmosphere}
\author[uva]{Justin Erwin}
\ead{jte2c@virginia.edu}
\author[um]{OJ Tucker}
\author[uva,nyu]{Robert E Johnson}

\address[uva]{Department of Engineering Physics, University of Virginia, 166 Engineers Way, P.O. Box 400745, Chalottesville, VA 22904}
\address[um]{Department of Atmospheric, Oceanic and Space Science, University of Michigan, Ann Arbor, MI 48109}
\address[nyu]{Department of Physics, New York University, New York, New York, 10012}

\begin{abstract}
Predicting the rate of escape and thermal structure of Pluto's upper atmosphere in preparation for the New Horizons Spacecraft encounter in 2015 is important for planning and interpreting the expected measurements. Having a moderate Jeans parameter Pluto's atmosphere does not fit the classic definition of Jeans escape for light species escaping from the terrestrial planets, nor does it fit the hydrodynamic outflow from comets and certain exoplanets. It has been proposed for some time that Pluto lies in the region of slow hydrodynamic escape. Using a hybrid fluid/molecular-kinetic model, we previously demonstrated the typical implementation of this model fails to correctly describe the appropriate temperature structure for the upper atmosphere for solar minimum conditions. Here we use a time-dependent solver to allow us to extend those simulations to higher heating rates and we examine fluid models in which Jeans-like escape expressions are used for the upper boundary conditions. We compare these to hybrid simulations of the atmosphere under heating conditions roughly representative of solar minimum and mean conditions as these bracket conditions expected during the New Horizon encounter. Although we find escape rates comparable to those previously estimated by the slow hydrodynamic escape model, and roughly consistent with energy limited escape, our model produces a much more extended atmosphere with higher temperatures roughly consistent with recent observations of CO. Such an extended atmosphere will be affected by Charon and will affect Pluto's interaction with the solar wind at the New Horizon encounter. Since we have previously shown that such models can be scaled, these results have implications for modeling exoplanet atmospheres for which the energy limited escape approximation is often used.
\end{abstract}


\maketitle

\section{Introduction}
Recent spacecraft exploration of the planets and moons in our solar system and the rapid increase in the discovery of exoplanets has increased interest in atmospheric escape from planetary bodies. The Cassini spacecraft is currently improving our understanding of  atmospheric escape from Titan, while in 2015 the New Horizons (NH) spacecraft will flyby Pluto and perform occultation observations of its and Charon's atmosphere \citep{guo2005}, and the Maven mission will orbit Mars studying the composition of its escaping atmosphere \citep{lin2012}. Furthermore, newly discovered exoplanets, with radii ranging from just a few times that of Earth to of the order of Jupiter, and their atmospheres are also modeled \citep{yelle2004, lammer2009, murrayclay2009}. Here we carry out simulations of Pluto's upper atmosphere, including atmospheric loss by thermal escape, that can be tested against data to be obtained during the NH encounter. By accurately describing the present loss rates, one can in principle learn about the evolution of Pluto's atmosphere. In addition, doing this accurately for a planet for which we will have in situ spacecraft data can, by scaling, guide our ability to model exoplanet atmospheres for which there will only be remote sensing data.

The previous models of atmospheric escape for Pluto used the concept of hydrodynamic escape by adapting the {\it critical solution} in \citet{parker1964a}. Parker described the expanding stellar corona and stellar wind assuming the temperature and pressure go to zero at infinity, and showed that for this to happen the bulk velocity must increase past the isothermal speed of sound at a critical point that is dependent on the temperature and gravity. 

This model was subsequently adapted for planetary atmospheres to include heating by solar radiation \citep{hunten1982, mcnutt1989} and was applied to Pluto \citep{kras1999, strobel2008} and Titan \citep{strobel2009}. It is often referred to as the slow hydrodynamic escape (SHE) model. The model requires solving the fluid equations out to very large distances from the planet to enforce the necessary boundary conditions. However, it is known that at some finite distance from the planetary surface the equations of fluid dynamics fail to describe the flow of mass, momentum and energy in an atmosphere \citep{johnson2010, volkov2011}. The region of validity of the fluid equation is often defined using the Knudsen number, $\Kn=\ell/H$, the ratio of the mean free path of the molecules, $\ell = (\sqrt{2}\sigma n)^{-1}$, to the density scale height, $H=\frac{k_b T}{m g(r)}$, of the atmosphere, where $\sigma$ is the collisional cross-section of the N$_2$ molecule, $n$ the local number density, $k_b$ the Boltzmann constant, $T$ the temperature, $m$ the mass of an N$_2$ molecule, and $g(r)$ the local gravitational acceleration. The fluid equations properly capture the physics where $\Kn\ll1$ so that many collisions occur over relevant length scales keeping the gas in local thermaldynamic equilibrium.

The concept of energy limited escape, which we will also examine here, is heavily used in modeling escape from early terrestrial planet atmospheres \citep{tian2009} and the growing body of data on exoplanet atmospheres \citep{lammer2009, valencia2010}. UV radiation drives escape by giving molecule the energy necessary to escape the gravitational well. Assuming that kinetic and thermal energy terms are small compared to gravity and thermal conduction is inefficient, the molecular loss rate is often approximated as
\begin{equation}
	\phi_L\approx\frac{Q}{U(r)}
\end{equation}
Here $Q$ is the EUV energy supplied in the upper atmosphere, and $U(r)=GMm/r$ is the gravitational energy with $G$ being the gravitational constant, $M$ the mass of Pluto, and $m$ the mass of an N$_2$ molecule. Without doing a detailed heating model one can use $Q=\pi r_{EUV}^2 \eta F_{EUV}$, where $r_{EUV}$ is the mean radius at which the radiation is absorbed and $\eta$ and $F_{EUV}$ are the heating efficiency and solar energy flux respectively. This depends critically on $r_{EUV}$ which is sometimes assumed to be close to the visual radius so that $r_{p}$ can be used. Also import is what level of the atmosphere the molecules are being removed from the gravitational well, again the visual radius is a good choice so that $U(r_p)$ is used.

The alternative to organized outflow is evaporative escape, by which the atmosphere loses gas on a molecule-by-molecule basis driven by conductive heat flow from below. The standard analytic model was originally developed by \citet{jeans} and is referred to as Jeans escape. The escape rates are found to depend predominately on the Jeans parameter $\lambda= U(r)/k_bT$, the ratio of gravitational to thermal energy of the molecules. For large Jeans parameters (i.e. when gravitational energy dominates thermal energy) the escape rate is obtained by integrating the velocity distribution over the portion of molecules that are moving upward with speed in excess of the escape velocity. Assuming a Maxwell-Boltzmann velocity distribution, the molecular escape rate $\phi$ and energy escape rate $\phi_E$ from an altitude $r$ are given by
\begin{align}
	\langle \phi \rangle_J &= \frac{1}{4}n\bar{v}\cdot(1+\lambda)\exp(-\lambda)\cdot 4 \pi r^2 \label{eq:jeans_n}\\
	\langle \phi_E \rangle_J &= \langle \phi \rangle_J \cdot k_bT\left(\frac{C_p}{k_b}-\frac{3}{2}+\frac{1}{1+\lambda}\right) \label{eq:jeans_e}
\end{align}
Here $\bar{v} =\sqrt{8k_bT/m\pi}$ is the mean molecular speed. To be consistent with the fluid equations, we consider expressions using a drifting Maxwell-Boltzmann to include the bulk velocity $u$ in the velocity distribution \citep{yelle2004, tian2009, volkov2011}. Typically these equations are applied at a level called the exobase where $\Kn\approx1$, as there are few collisions above this level to inhibit a molecule from escaping. 

The Jeans escape boundary condition has been used in modeling planetary atmospheres, notably by \citet{chassefiere1996}, \citet{yelle2004} and \citet{tian2008}. In modeling the response of the Earth's thermosphere to EUV heating, \citeauthor{tian2008} used the Jeans escape expression to determine the velocity at the upper boundary (the exobase in their case); however at the upper boundary a zero temperature gradient was applied to neutrals, and a fixed heat gradient was applied to electrons separately. In modeling extra-solar gas giants at small orbital distance, \citeauthor{yelle2004} also used Jeans escape, with a drifting Maxwell-Boltzmann velocity distribution, to get an upper boundary velocity. However, this was applied at a fixed upper boundary of 3 planetary radii rather than at the exobase.

Accurate calculations of thermal escape can be obtained, in principle, by solving the Boltzmann equation or, more often, by using the direct simulated Monte Carlo (DSMC) method as in \citet{tucker2009} and \citet{volkov2011}. In this method, representative particles are used along with a model of molecular collisions to calculate the temperature, density and other macroscopic values from the velocity distribution in both the dense and rarified regions. In this respect, DSMC can model the exospheres with better accuracy than hydrodynamic model, but can only model the lower, denser regions of the atmosphere at great computational cost. Using the DMSC method \citet{volkov2011} demonstrated earlier that for a monatomic or diatomic gas in the absence of heating above some lower boundary for $\lambda\gtrsim3$ the molecular escape rate is Jeans-like at the exobase (i.e. $\phi/\langle \phi \rangle_J\approx$ 1.5 for the range of Jeans parameters studied). For our model of Pluto, the lower boundary of the simulated domain has $\lambda=22.8$, and the exobase values exceed 4. Using the above results as guidance, we model the principle component in Pluto's atmosphere to obtain a description relevant to the NH encounter. We did this in order to test energy-limited escape, and to better understand the transition from Jeans to hydrodynamic escape.

\section{Model Description}
The steady state equations of mass, momentum, and energy have been used to study hydrodynamic escape. They can be solved to give the radial dependence of number density $n$, outward bulk velocity $u$, and temperature $T$ in the region in which the atmosphere is collisional. However, as we have shown \citep{tucker2012, volkov2011} they cannot be used by themselves to determine the escape rate unless the Jeans parameter is very small, in which case the solutions are somewhat insensitive to the boundary condition at infinity. The equations, neglecting viscosity can be written as
\begin{align}
	4 \pi r^2 n u = \phi \label{eq:fluid_mass}\\
	nm\frac{\partial}{\partial r}\left({\textstyle\frac{1}{2}}u^2\right) + \frac{\partial p}{\partial r} = -nmg(r)
		 \label{eq:fluid_momentum} \\
	{ \frac{\partial}{\partial r}\left( \phi\left(C_pT+{\textstyle \frac{1}{2}}mu^2-U(r)\right) - 
	 	4\pi r^2 \kappa(T)\frac{\partial T}{\partial r} \right) } \nonumber \\
		= 4 \pi r^2 q(r) 
		\label{eq:fluid_energy}
\end{align}
Here $\phi$ is the molecular escape rate through the one dimensional atmosphere, $p$ is the pressure (related via the equation of state, $p=nk_bT$), $m$ is the mass of a N$_2$ molecule, $\kappa(T)$ is the conductivity, $C_p=\frac{7}{2}k_b$ is the specific heat at constant pressure, and $q(r)$ is the net heating/cooling rate per unit volume. In Eq. (\ref{eq:fluid_energy}), the first term on the left side is work done by adiabatic expansion and the second is the work done by conduction.

For the conductivity, we use the power law $\kappa(T)=\kappa_0T^\omega$ to approximate the temperature dependence. Some authors use an empirical fit for the conductivity, e.g. \citet{hunten1982} used $\omega = 1.12$, while \citet{mcnutt1989} used $\omega=1$ since it simplifies the analytic solutions to Eq.~(\ref{eq:fluid_energy}). In this paper we use $\kappa_0=9.37\times10^{-5}$ J/m K and $\omega=1$ to compare with \citet{strobel2008} and because this is consistent with the variable hard sphere model for collisions between N$_2$ molecules used in the DSMC model of the exosphere \citep{tucker2012, volkov2011}.

The lower boundary of our fluid domain is set at 1450 km, consistent with the occultation results and the assumptions of \citet{strobel2008}.  As we will show, we need to simulate at least up to the exobase so that we can determine an escape rate and enforce our upper boundary condition on temperature. Since we showed earlier \citet{tucker2012} that the atmosphere is highly extended, the upper boundary is set to 16000 km to accommodate the solar maximum case. The radial grid requires fine spatial resolution for the first 3000 km where the bulk of the heating occurs; however, a coarse grid is sufficient above this. Therefore, we use a grid that is equally spaced in $1/r$. With $N_r=1600$ radial grid points in $r$, the lower boundary $\Delta r$ of order 1 km and the upper boundary $\Delta r$ of order 100 km.

Previously, we directly solved the steady state Eqs.~(\ref{eq:fluid_mass}-\ref{eq:fluid_energy}) by iteratively solving them along with the heating and the DSMC escape simulations until a consistent solution was found \citep{tucker2012}. Here we reintroduce the time-dependence into the energy equation and iterate through time until convergence is achieved. This is found to be more numerically stable for the boundary conditions and the higher heating cases that we will consider here. We make the substitution $\xi=T^{\omega+1}$ as in \citep{strobel1996, zalucha2011} in the time-dependent form of Eq.~(\ref{eq:fluid_energy}) to obtain a linear time-dependent PDE:
\begin{equation}
	\begin{aligned}
 	\frac{d\xi}{dt} =& \frac{(\omega+1)T^\omega}{n C_p}\left( q(r) - \frac{\phi}{4\pi r^2}\frac{d}{dr}\left(\frac{1}{2}mu^2-U(r)\right) \right) \\
 	&+ \left(\frac{\kappa_0 T^\omega}{n C_p}\frac{2}{r}-\frac{\phi}{4\pi r^2n}\right)\frac{\partial\xi}{\partial r}
 		+ \frac{\kappa_0 T^\omega}{n C_p}\frac{\partial^2\xi}{\partial r^2} 
	\end{aligned}
\label{eq:energy2}
\end{equation}
This equation is solved using the implicit, finite difference time stepping scheme described in Appendix A. This second-order equation needs two boundary conditions. The first is the fixed lower boundary temperature $T(r_0) = 88.2$K, and the second is a restriction on the energy flux leaving the atmosphere from the top boundary $r_{top}$
\begin{equation}
	4\pi r^2 \frac{\kappa_0}{\omega+1}\left.\frac{d\xi}{dr}\right|_{r_{top}} = \phi\left(C_pT+\frac{1}{2}mu^2-U(r)\right)_{r_{top}} - \phi_E
	\label{eq:ubc}
\end{equation}
This condition is derived by integrating Eq.~(\ref{eq:fluid_energy}) from the exobase to the upper boundary and the energy flux from the atmosphere, $\phi_E$, is determined by the escape model, the details of which will be explained in Section (\ref{sec:hybridmodels})

The time stepping advances the temperature alone. To update density and bulk velocity, we first update the values of $u$ using Eq.~(\ref{eq:fluid_mass}) and the previous values of $n$. This new values of $u$ are then used to update $n$ using the momentum equation (\ref{eq:fluid_momentum}) re-expressed as
\begin{equation}
	p(r) = p_0\exp\left(-\int_{r_0}^r{\frac{\frac{1}{2}m\frac{d(u^2)}{dr}+mg(r)}{k_bT}dr}\right) \label{eq:fluid_mom_int}
\end{equation}
As Eqs.~(\ref{eq:fluid_mass}) and (\ref{eq:fluid_mom_int}) are coupled through $n$ and $u$, we iteratively solve them together to find a consistent solution. A few iterations are sufficient to converge on a self-consistent solution for $T$, $n$, and $u$. With the profiles updated, the last action of the time step is to recalculate the heating rates.

The radiative heating and cooling model is adapted from \citet{strobel2008}, using the same fixed mixing ratios of $\chi_{N_2}=0.97$, $\chi_{CH_4}=0.03$, and $\chi_{CO}=0.00046$ for computing the heating and cooling rates. The energy fluxes and effective cross section for FUV absorption by CH$_4$ and EUV absorption by N$_2$ are given in \citet{kras1999} for the various levels of solar activity at Pluto's perihelion (i.e. 30 AU). We obtain globally averaged heating rates for the FUV/EUV by applying Lambert-Beer's Law for a plane-parallel atmosphere separately for each species $s$. Assuming an incoming energy flux $F_{s}^{\infty}$, the energy flux at a given altitude is given by
\begin{equation}
	F_s(r) = \frac{1}{2} \mu F_{s}^{\infty} \exp(-\tau_s(r)/\mu)
\end{equation}
where $\tau_s(r)=\int_r^{\infty}\sigma_s n_s(r) dr$ is the vertical optical depth, $\sigma_s$ is the absorption cross section of species $s$, $n_s$ is the number density of species $s$, and $\mu=\cos(60^\circ) = 0.5$ is used to approximate spherically averaged heating \citep{apruzese1980, strobel1996}. Then the heating rate for each species is given by
\begin{equation}
	q_s(r) = \frac{dF_s}{dr} = \sigma_s n_s(r)F_s^\infty \exp(-\tau_s(r)/\mu)
\end{equation}
Further heating and cooling mechanisms are provided by methane near-IR absorption and CO rotational line emission. In \citet{strobel1996} the methane near-IR heating was treated as an non-LTE process, and the CO rotational line emission was shown to be an LTE process for pressure greater than $10^{-5}$ Pa while above this it may become an non-LTE process. Parameterizations of these processes are given in \citet{strobel2008}, fit to the detailed radiative-transfer model of Pluto's lower atmosphere of \citet{strobel1996}. The methane heating rate per molecule is given as $8\times10^{-22}\exp(-\frac{\lambda_0-\lambda}{1.6})$ erg s$^{-1}$, where $\lambda=\frac{GMm}{rk_bT_0}$ and $\lambda_0 = \lambda(r_0)$. The CO rotational cooling rate per molecule is given as $8\times10^{-20}(\frac{T}{105})^{2.45}$ erg s$^{-1}$. 

These calculations assume a fixed efficiency for the various processes, yet we know that this assumption will break down when the gas becomes tenuous. To accommodate this, we assume the heating rate is zero (i.e. the heating efficiency drops to zero) above $\Kn=0.1$. Furthermore, with the heating per molecule constant, including heating up to the top boundary can cause numerical instabilities even though the heating rate per unit volume is small. In describing the heating, a cut-off is often used where the tangential line-of-sight optical depth is unity \citep{strobel2008}, which in our case lies below the $\Kn=0.1$ level.

To aid in convergence to steady state, we estimate the heating $q$ at the advanced time step using a Taylor expansion in $\xi$ as $q(r,\xi^{i+1}) = q(r,\xi^i) + \frac{dq}{d\xi}(\xi^{i+1}-\xi^i)$. This can help dissipate numerical oscillations between time steps, speeding up convergence. In calculating the derivative for the FUV and EUV heating we assume $\frac{dn}{dT}=0$ and $\frac{d\tau}{dT} = \frac{\tau}{T}$ as the optical depth $\tau \approx \sigma n H$ with $H$ depending linearly on $T$.

\subsection{Hybrid Models} \label{sec:hybridmodels}

The hybrid model described here entail finding a fluid model that is consistent with a kinetic model of escape, as opposed to homogeneous boundary condition at infinity of the hydrodynamic models. The two models are coupled since the kinetic model determines the escape rate $\phi$ and energy escape rate $\phi_E$, but depends on the solution of the fluid model. While the temperature found through Eq.~(\ref{eq:energy2}) depends on $\phi$ with the upper boundary condition in Eq.~(\ref{eq:ubc}) also depending on $\phi_E$. Furthermore, density and bulk velocity found through the coupled Eqs.~(\ref{eq:fluid_mass}) and (\ref{eq:fluid_mom_int}) depend on $\phi$ and the temperature structure. In this paper we describe two hybrid models of escape:

(i) The first hybrid models we use is to combine the fluid model of the thermosphere with a DSMC model of the exosphere as was done in \citet{tucker2012}. This we refer to as the fluid-DSMC model. We refer readers to \citet{tucker2012} for the details of the DSMC model used here, with the only change from that paper is to use the time stepping method in the fluid model instead of the steady state fluid solution. This has been found to converge better in the presence of high heating, giving us the ability to find a fluid-DSMC solution for solar mean condition that is describe in the results section below.

The fluid equations and the DSMC simulation are coupled as the fluid equations depend on the escape rate $\phi$ and energy escape rate $\phi_E$ obtained from the DSMC simulation, while the DSMC simulation depends on the choice of temperature $T$, density $n$, and location $r$ of its lower boundary. The fluid-DSMC solution is the self consistent solution obtained through iteration.

We begin with an estimate of $\phi$ and $\phi_E$, and time step the fluid equation until convergence to obtain our first fluid solution. We locate where $\Kn = 0.1$ of the fluid solution, and use the value of $T$, $n$, and $r$ to begin a DSMC simulation. This is far enough below the exobase, in the collision-dominated regime, to act as a lower boundary for the single component DSMC simulation, and above which heating can be ignored based upon the previous assumption of zero heating efficiency. The DSMC results for escape $\phi$ and $\phi_E$ are then used in the next fluid solution. After a few exchanges between the fluid and DSMC models, a consistent solutions is found for each specific heating case (we have solutions for no heating, solar minimum, and solar mean heating). Although non-thermal processes in the exobase region contribute to and can dominate the atmospheric loss rate (e.g. photo-disassociation, atmospheric sputtering), we ignore these here, consistent with previous work on escape from Pluto. Such processes will be included in subsequent work.

(ii) In the second hybrid model we find a fluid solution that is consistent with Jeans escape from the exobase, hereafter referred to as the fluid-Jeans model. In each iteration we find the exobase, $\Kn=1$, and update the escape rate and energy escape rate based on the Jeans equations (\ref{eq:jeans_n}-\ref{eq:jeans_e}) by using $\phi = K\times\langle\phi\rangle_J$ and $\phi_E = K\times\langle\phi_E\rangle_J$. Using $K=1$ results is normal Jeans escape, while $K>1$ is used to better reflect the results of molecular kinetic modeling that shows Jeans escape consistently underestimates the rate for the range of $\lambda$ of interest here \citep{merryfield1994, volkov2011, tucker2012} . A consistent solution is obtained by stepping in time until convergence, which we define as when the 2-norm of the change in temperature, $\Delta T_i$, multiplied by the time step size divided by the number of radial grid points, $\left(\sum \Delta T_i^2 \right)^{1/2}\Delta t/N_{r}$, is less than $10^{-5}$ K s$^{-1}$.

The zero heating case was initiated using an isothermal profile and time stepped to steady state. Then we incrementally increased the solar flux rates to obtain the other 3 profiles in turn. As the exobase altitude increased with increase in the heating rate, the upper boundary altitude had to be increased to 16000 km (or about 14$r_p$) to model the solar maximum condition.

The fluid-DSMC model is the more physically accurate model since it correctly captures the transition from collisional to near collisionless flow in the atmosphere. But the DSMC portion of the iteration requires significant computational expense and is not practical for parameter exploration. The fluid-Jeans model is used here to quickly perform parameter studies and to extrapolate to the higher heating rates. Estimates of the enhancement in the escape rate relative to JeanÕs escape rate can be used to roughly approximate the atmospheric loss rate and as a starting point for a full fluid-DSMC simulation.


\section{Results}
\subsection{Fluid-DSMC}
\begin{figure*}[p]
\centering
 \includegraphics[width = 0.7\textwidth]{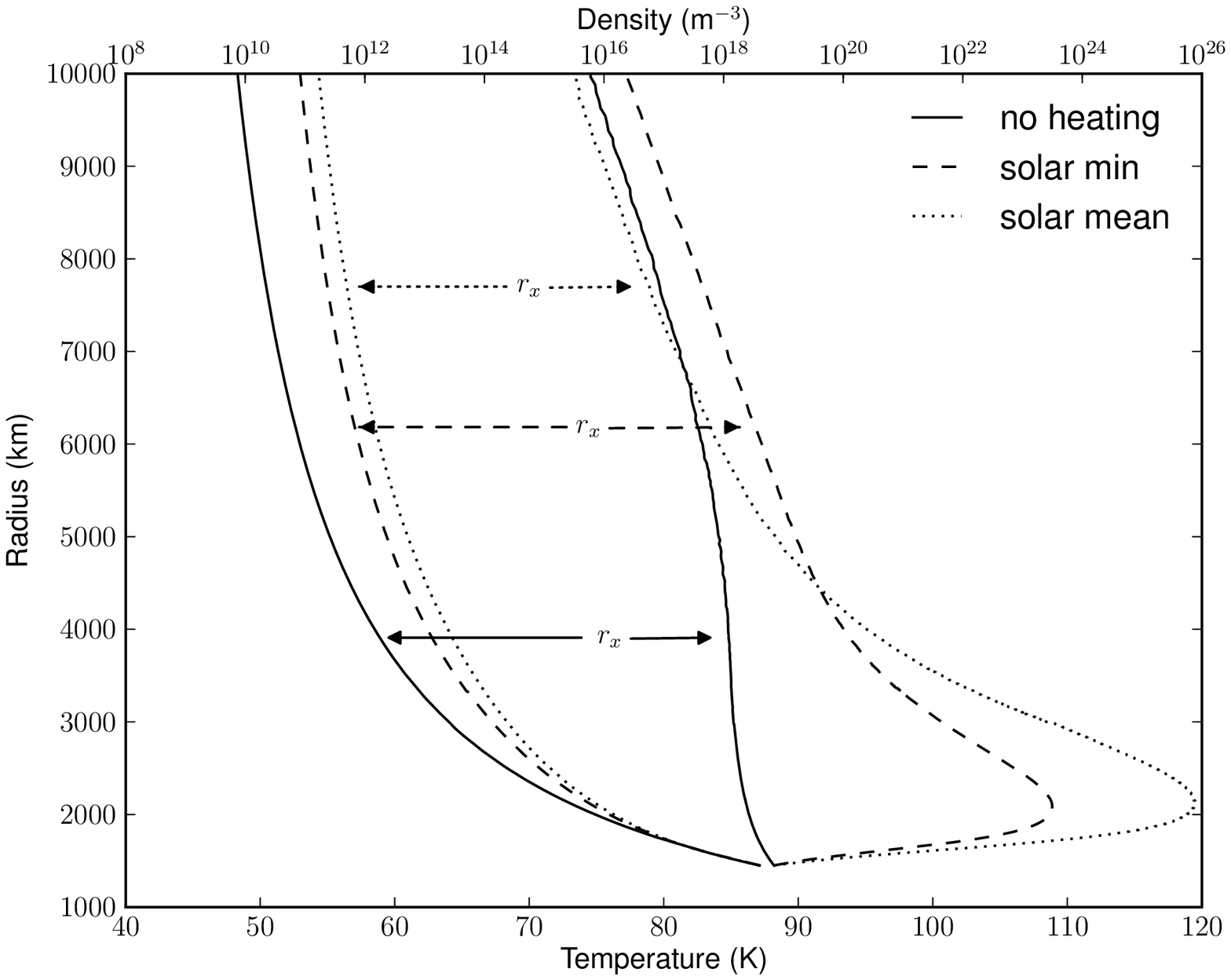}
 \caption{Temperature (right) and density (left) profiles for the hybrid fluid-DSMC model of Pluto, for the cases of zero heating and solar minimum \citep[from][]{tucker2012}, and our new result for solar mean. The temperate profiles are plotted with the scale labeled below; the density profiles are plotted on a logarithmic scale, with the scale labeled above. The exobase altitudes are marked for each case by $r_x$. The DSMC solutions are time-averaged after reaching steady state to smooth out random oscillations.}
 \label{fig:fdsmc_Tn}
 \includegraphics[width = 0.7\textwidth]{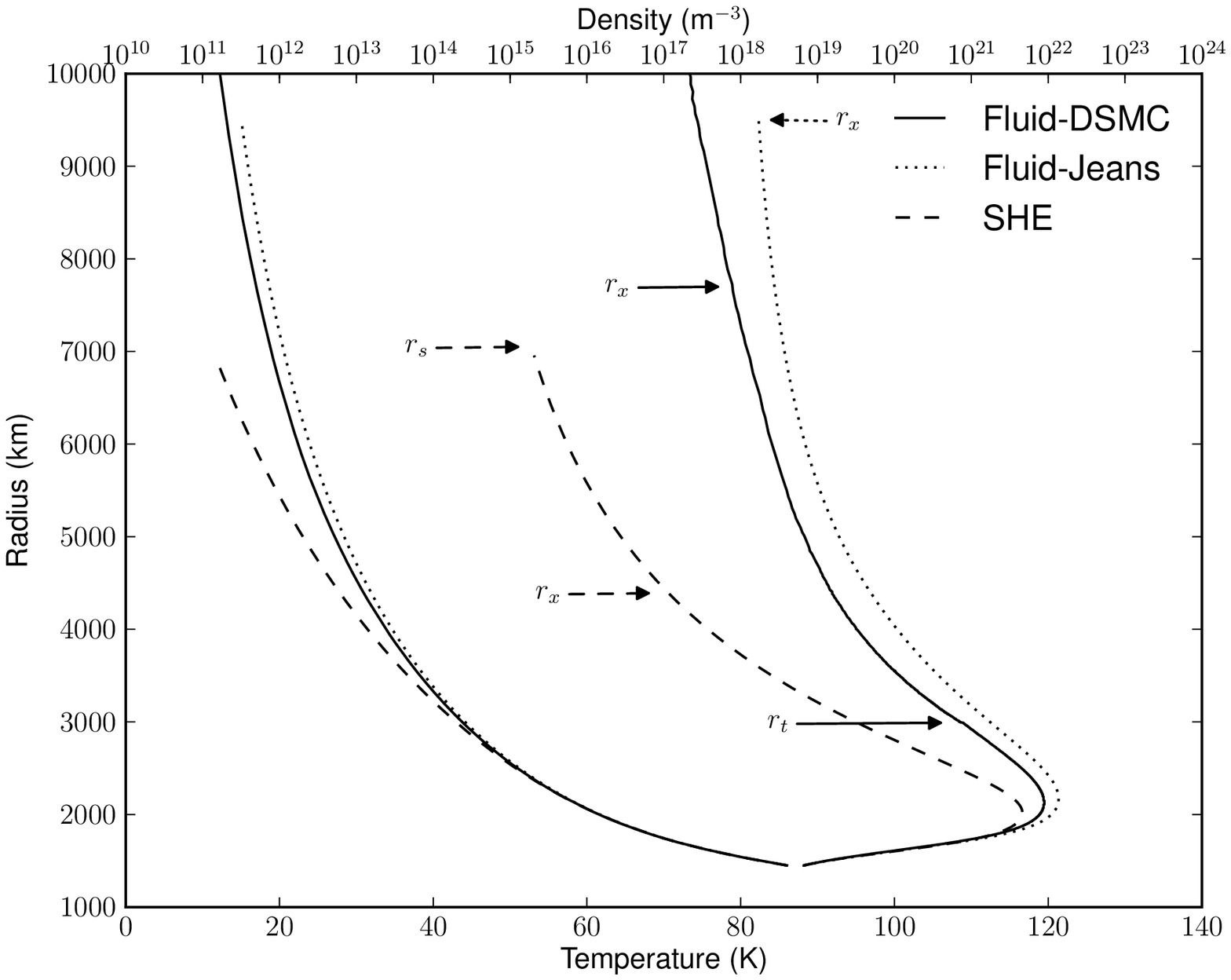} 
 \caption{Temperature (right) and density (left) profiles for solar mean conditions, comparing the fluid-DSMC solution of the present study with that of the SHE from \citet{strobel2008}, along with the fluid-Jeans solution describe in Section~\ref{subsec:res:fjeans}. Marked on the fluid-DSMC solution is the exobase ($r_x$) as well as the transition between the fluid model and DSMC ($r_t$, where $\Kn(r_t)=0.1$). Marked on the SHE model are the exobase ($r_x$) and the sonic point ($r_s$).}
 \label{fig:temp_comparesolarmed}
\end{figure*}

With the time-dependent solver, which is an improvement on the steady state iterations used in \cite{tucker2012}, we are able to model higher solar heating rates up to and including solar mean. In Figure~\ref{fig:fdsmc_Tn} we present the two previously calculated solutions found in that paper for zero heating and solar minimum heating along with a new fluid-DSMC hybrid result for solar mean heating. This allows us to bracket the expected heating rates for the NH encounter. The hybrid solution includes rotational/ translational energy exchange consistent with the assumed thermal conductivity \citep[see][]{tucker2012}. Vibrational energy exchange of N$_2$ is not included since the characteristic temperature is not in excess of 3000K \citep{bird}.

A few qualitative differences between the hydrodynamic solutions and our fluid-DMSC solutions are seen in Figure~\ref{fig:temp_comparesolarmed}. Although the two types of calculations give similar escape rates, they produce different temperature profiles also discussed in \citet{volkov2011} and  \citet{tucker2012}. More importantly for the NH encounter, the exobase altitude for the fluid-DSMC solutions are much higher (6000-10000km depending on the heating rate) suggesting a significantly more extended atmosphere for Pluto than suggested by previous simulations. This extended atmosphere may support recent observation of CO out to as far out as 4500km \citep{greaves2011}. Although, the hydrodynamic methods can produce atmospheres that expand at supersonic speeds, we find the expansion is subsonic with the bulk flow velocity being less than 3\% the speed of sound at the exobase (the speed of sound is $\sim$180 m s$^{-1}$ depending on temperature). Further out in the exosphere, the bulk speed increases to approach a constant value well below the speed of sound, consistent with a Jeans-like escape model.

In the solar mean case using the fluid-DSMC model, the transition from the fluid to DSMC domain is near 3000 km. In the region $0.1<\Kn<0.2$, the agreement between the fluid and DSMC solutions is better than 5\% as in \citet{tucker2012}. The DSMC solution begins to deviate from the fluid solution above the $\Kn=0.2$ level. Here the kinetic and internal energies of the molecules begin to separate, which is associated with the transition to non-equilibrium flow. It is important to note that this occurs below the exobase, so the region of validity of the fluid equations should be terminated below the exobase. Our choice of $\Kn = 0.1$ for connecting the DSMC solution to the fluid solution is at an altitude below which this separation occurs.

The solutions of \citet{strobel2008} result in exobase altitudes much lower than that found in our fluid-DSMC model. His exobase altitudes are $\sim$0.56 - 0.7 of those calculated with the hybrid model. With the lower exobase altitudes and the lower temperatures computed by \citet{strobel2008}, the Jeans parameter at the exobase, $\lambda_x$, is $\sim$2.6 - 3.3 times larger than found in this study. Using such a high $\lambda_x$ in Eq.~(\ref{eq:jeans_n}) leads to a Jeans escape rate is significantly lower than the total escape rate found; hence the earlier conclusion that escape rates were orders of magnitude larger than the Jeans escape rate. In contrast, using the exobase temperature calculate using the fluid-DSMC model results in an escape rate that is only modestly enhanced relative to the calculated Jeans escape rate, the ratio $\phi/\langle\phi\rangle_J$ being 1.6, 2.0 and 2.3 for zero, solar minimum, and solar mean, respectively, as seen in Table~\ref{tab:fdsmc}.

\begin{table}[t]
 \centering
 \begin{tabular}{lccc}
 	\hline\hline
 	Solar Heating									& zero		& min		& mean 	\\ \hline
	$Q$ ($10^{14}$ erg s$^{-1}$)			& 0			& 3.8		& 7.8		\\
	$\phi$ ($10^{27}$ s$^{-1}$)				& 0.047	& 1.20		& 2.56		\\
	$E_{esc}$	 ($10^{-3}$ eV)			& 14.6		& 14.3		& 14.3		\\
	$r_x/r_p$											& 3.37		& 5.24		& 6.71		\\
	$\lambda_x$									& 8.9		& 5.7		& 4.8		\\
	$\phi/\langle\phi\rangle_J$				& 1.6		& 2.0		& 2.3		\\
	$\phi/\phi_L$									& -			& .88		& .91		\\
	$E_{esc}/\langle E_{esc}\rangle_J$	& .95	& .90		& .97		\\ [0.5ex]
 	\hline
 \end{tabular}
 \caption{Fluid-DSMC results. $E_{esc} = \phi_E/\phi$ is the average energy carried off per molecule; $r_x$ is given as a ratio to $r_p=1150$ km. $\langle\phi\rangle_J$ is computed using exobase values from the solution, $\phi_L =Q/U(r)$ is computed using $Q$ from the solution and $r=1450$ km.}
 \label{tab:fdsmc}
\end{table}

The results in Figure~\ref{fig:fdsmc_Tn} and Table~\ref{tab:fdsmc} correspond to a net solar heating/cooling $Q=\int_{r_0}^{r_x}{4\pi r^2 q(r) dr}$ of 0.0, 3.8, and 7.8 $\times 10^{14}$ erg s$^{-1}$ for zero, solar minimum, and solar mean, respectively. These can be compared to 0.0, 3.4, and 6.9 $\times 10^{14}$ erg s$^{-1}$ for zero, solar minimum, and solar mean, respectively, found in \citet{strobel2008}. To give context for these heating rates, the upward flow of energy across Pluto's 1450 km level would be greater than $10^{20}$ erg s$^{-1}$ if it were escaping hydrodynamically (i.e. $\lambda \rightarrow 0$).

Many of the relevant values of the fluid-DSMC hybrid solution in \citet{tucker2012}, as well as those for the new solar mean heating case, can be found in Table~\ref{tab:fdsmc}.  It is seen that the escape fluxes are comparable to the Jeans flux and are also, quite remarkably, comparable to our estimate of the energy limited rate with the exception, of course, of the $Q=0$ case. This result alone is important because it is often presumed that the Jeans escape rate and the energy-limited rate are two limiting cases for atmospheric loss.

As in \citet{tucker2012}, for zero heating our escape rate is a order of magnitude lower than the SHE solution. For solar minimum our total escape rate is 20\% below the SHE solution, and for solar mean we are slightly above. In each of these simulation cases, somewhat more heat is deposited than in \citet{strobel2008} due to our higher density and highly extended atmosphere. If the current simulation were scaled to the same heating rates the fluid-DSMC escape rate would be slightly smaller than the SHE model as the DSMC simulations produce higher temperatures in the exosphere region, which imply lower adiabatic cooling and hence lower escape rate.

\subsection{Fluid-Jeans}
\label{subsec:res:fjeans}
The fluid-DSMC solutions demonstrate that the escape rate and energy escape rate are within a factor of a few times the Jeans escape rate and energy escape rate as calculated from Eqs.~(\ref{eq:jeans_n}) and (\ref{eq:jeans_e}). Therefore, in Figure~\ref{fig:temp_comparesolarmed} and Table~\ref{tab:fjeans} we present results for the Fluid-Jeans solution using an upper boundary condition that is equal the Jeans escape and associated energy loss rates, and then doubling those. We note that average energy carried off per molecule in the Fluid-DSMC simulations is near that predicted by Jeans escape. By scaling both the molecular and energy escape rates by the same constant, the energy per molecule is not scaled. In Figure~\ref{fig:temp_comparesolarmed}, it is seen that the fluid-Jeans solution is in many respects similar to that of fluid-DSMC solution (i.e. high temperature in upper thermosphere, extended exobase, etc.). Therefore, approximating the escape process as Jeans-like for Pluto's atmosphere under solar heating conditions likely occurring at the NH encounter reproduces many of the properties of the full solution in which we are interested. Although the fluid equations may not be valid up to the exobase, using them along with Jeans escape boundary conditions allows us to carry out parameter studies that are impractical due to computational cost with the full fluid-DSMC model.

\begin{table}[ht]
 \centering
 \begin{tabular}{lccc}
 	\hline\hline
	\multicolumn{4}{c}{Using $1\times$ Jeans Escape Rate} \\
 														& zero		& min		& mean 	\\ \hline
	$Q$ ($10^{14}$ erg s$^{-1}$)		& 0			& 3.86		& 7.87	\\
	$\phi$ ($10^{27}$ s$^{-1}$)			& 0.035	& 1.14		& 2.58	\\
	$E_{esc}$	 ($10^{-3}$ eV)		& 15.5		& 16.3		& 14.6	\\
	$r_x/r_p$										& 3.41		& 5.99		& 8.10	\\
	$\lambda_x$								& 8.71		& 4.72		& 3.84	\\ [0.5ex]
	\hline\hline
	\multicolumn{4}{c}{Using $2\times$ Jeans Escape Rate} \\
 														& zero		& min		& mean 	\\ \hline
	$Q$ ($10^{14}$ erg s$^{-1}$)		& 0			& 3.83		& 7.79	\\
	$\phi$ ($10^{27}$ s$^{-1}$)			& 0.054	& 1.17		& 2.58	\\
	$E_{esc}$	 ($10^{-3}$ eV)		& 15.2		& 15.4		& 13.7	\\
	$r_x/r_p$										& 3.34		& 5.30		& 6.84	\\
	$\lambda_x$								& 9.02		& 5.55		& 4.72	\\ [0.5ex]
 	\hline
 \end{tabular}
 \caption{Results of the Fluid-Jeans model. The first results uses Jeans escape rate values for $\phi$ and $\phi_E$, while the second set uses twice the Jeans escape rate values. The exobase radius, $r_x$, is given a ratio relative to $r_p=1150$ km.}
 \label{tab:fjeans}
\end{table}

\begin{figure*}[ht]
	\centering
 	\includegraphics[width = 0.7\textwidth]{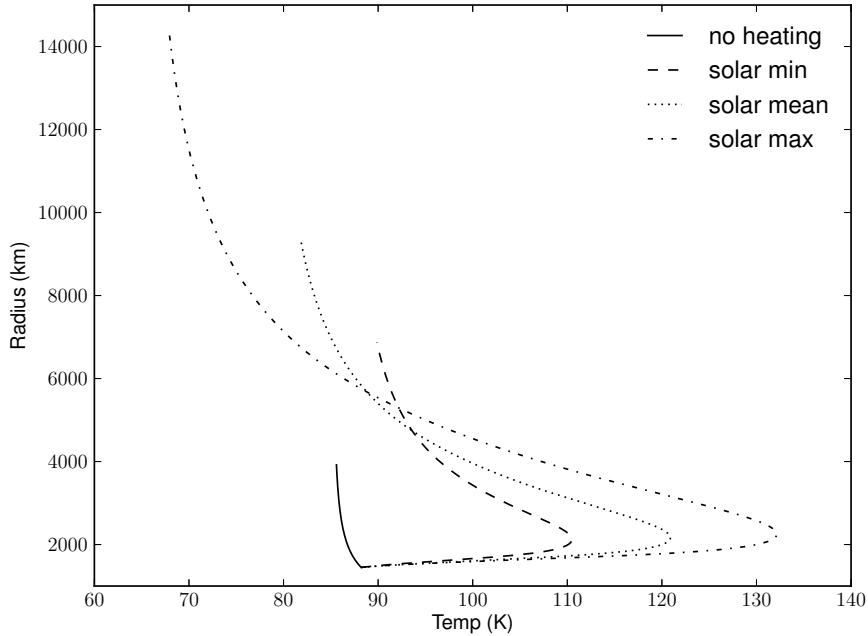}
 	\caption{Temperature profile found in the fluid-Jeans model for the 4 special cases discussed: zero heating, solar minimum, solar mean, and solar maximum. The top of each curve represents the exobase altitude.}
 	\label{fig:fjeans_temp}
\end{figure*}

The Fluid-Jeans model using the Jeans escape rate from the exobase results in a total escape rate is consistent with the fluid-DSMC model, but several properties of the fluid-Jeans solution differ from the fluid-DMSC solution. For instance, the atmosphere is more extended than in the fluid-DSMC model, causing slightly more EUV heat to be absorbed.. The Jeans parameter at the exobase is lower than for fluid-DSMC model. These differences increase with escape rate. 

To more accurately model the escape, also presented in Table~\ref{tab:fjeans} are results obtained by scaling the Jeans escape values by a factor of 2 for the upper boundary, approximating the enhancement seen in the fluid-DSMC solutions in Table~\ref{tab:fdsmc}. With this enhancement, the escape rate, exobase altitude, and exobase Jeans parameter more closely resemble the fluid-DSMC results in Table~\ref{tab:fdsmc}. Surprisingly, ignoring the zero heating case, the converged escape rate is barely affected by the enhancement of the Jeans boundary conditions. Similarly, increasing the enhancement to 3 and 4 times the Jeans escape values increases the escape rate only a few percent, while the exobase altitude and temperatures in the upper atmosphere decrease significantly. Since the escape rate is roughly given by the energy-limited approximation, this is consistent with the loss rates being insensitive to the size of the enhancement in the Jeans boundary conditions. Therefore, the other atmospheric properties adjust to ensure a nearly constant escape rate. However, we would not have had an estimate of the enhancement of the escape rate relative to Jeans to accurately predict the atmospheric structure in the absence of the fluid-DSMC simulations. Our early DSMC simulations indicated that for the case of zero heating above the lower boundary of the simulation region, the enhancement in total escape rate is $\sim$1.6 \citep{volkov2011}. How the escape rate enhancement increases with heating remains an open problem. 

As the heating rate increases, the atmosphere gets further extended and $\lambda_x$  decreases to  $\sim$3 at the exobase for the solar maximum case using no enhancement to the total escape rate relative to Jeans escape rate. This is within the region that \citet{volkov2011} found marks the transition to hydrodynamic escape for N$_2$ at the relevant temperatures. In that study, it was shown that for the lower boundary at small $\Kn$, the exobase moves out to infinity as the Jeans parameter approaches 3. Using twice the Jeans escape values from Eqs.~(\ref{eq:jeans_n}) and (\ref{eq:jeans_e}) as the boundary conditions causes the Jeans parameter in the upper atmosphere to increase, keeping the solution from entering the transonic regime. 

\begin{table}[ht]
 	\centering
 	\begin{tabular}{lccc}
 		\hline\hline
		\multicolumn{4}{c}{solar maximum} \\
 		Jeans escape rate -					& $1\times$		& $2\times$		& $3\times$ 	\\ \hline
		$Q$ ($10^{14}$ erg s$^{-1}$)		& 15.8				& 15.7				& 15.6	\\
		$\phi$ ($10^{27}$ s$^{-1}$)			& 5.81				& 5.80				& 5.79	\\
		$E_{esc}$	 ($10^{-3}$ eV)		& 11.7				& 10.9				& 10.4	\\
		$r_x/r_p$										& 12.4				& 9.8				& 8.9	\\
		$\lambda_x$								& 3.01				& 3.95				& 4.48	\\ [0.5ex]
		\hline
 	\end{tabular}
 	\caption{fluid-Jeans model results for solar maximum case. Results are given for $1\times$, $2\times$, and $3\times$ the Jeans escape values as an upper boundary condition. The last column (3$\times$ Jeans) is an estimate of the enhancement, inferred from the fluid-DSMC results in Table 1.}
	\label{tab:fjeans_max}
\end{table}

Based on the escape rate enhancement as a function of heating rate found in Table~\ref{tab:fdsmc}, the escape rate for solar maximum condition is estimated to be close to 3 times the Jeans escape rate. If we accept the assumption that the enhancement will be 3.0 times Jeans for the solar maximum case then we can use the fluid-Jeans model to discuss the atmosphere. In Table~\ref{tab:fjeans_max} the heating rate has approximately doubled relative to solar mean and the escape rate has likewise approximately doubled. The average energy carried off per molecule has decreased to $\sim11\times$10$^{-3}$eV, a decrease $\sim$30\% relative to solar mean. The fluid-Jeans solutions consistently underestimate the energy carried off per molecule, so the values given are likely lower than the values a fluid-DSMC solution would give. The energy-limited escape rate $\phi_L=5.61\times10^{27}$ for this $Q$ is close to our estimated escape rate.

In Figure~\ref{fig:fjeans_temp}, the results for the four cases considered here (using the Jeans criteria in Eqs.~(\ref{eq:jeans_n}) and (\ref{eq:jeans_e}) at the upper boundary) are shown. We see that unlike the exobase altitude (that varies from $\sim$4000 to 14000km), the altitude of the peak translational temperature does not increase significantly with heating rate, changing from 2124 km to 2250 km from solar minimum to solar max with translational temperatures of 110K and 132K, respectively. The altitude of the heating peak is below that of the temperature maximum, and also does not change significantly with heating rate, changing from 1767 km to 1780 km from solar minimum to solar maximum, respectively. The altitude of the heating peak is near where the CH$_4$ optical depth in the FUV is unity. The fluid-Jeans solutions become nearly isothermal at the exobase consistent with the fluid-DSMC model, but differ from the \citet{parker1964a} and SHE models \citep{mcnutt1989, watson1981, kras1999, strobel2008} that require the temperature to decrease to zero at infinity.

\cite{tian2008} found two regimes of escape in response to increased UV heating in his fluid dynamic studies of an early Earth atmosphere using the Jeans escape rate for the upper boundary condition. The transition from hydrostatic equilibrium to what he calls hydrodynamic escape coincides with adiabatic cooling becoming non-neg\-ligible when compared to conduction or radiative heating. In our simulations of Pluto's atmosphere, adiabatic cooling is always an important cooling mechanism in the upper atmosphere. Furthermore, the changes in exobase altitude, temperature and bulk velocity with increased heating are similar to hydrodynamic escape regime of \citet{tian2008}, even though our escape is described as a kinetic process. Adiabatic cooling is always important in our simulations likely due to the much smaller gravitational binding energy of Pluto compared to Earth. The similarities with \citet{tian2008} mentioned are likely due to his use of the Jeans escape rate at the upper boundary, which we have shown can be a reasonable approximation to the kinetic boundary conditions.

\begin{figure*}[t]
	\centering
 	\includegraphics[width = 0.49\textwidth]{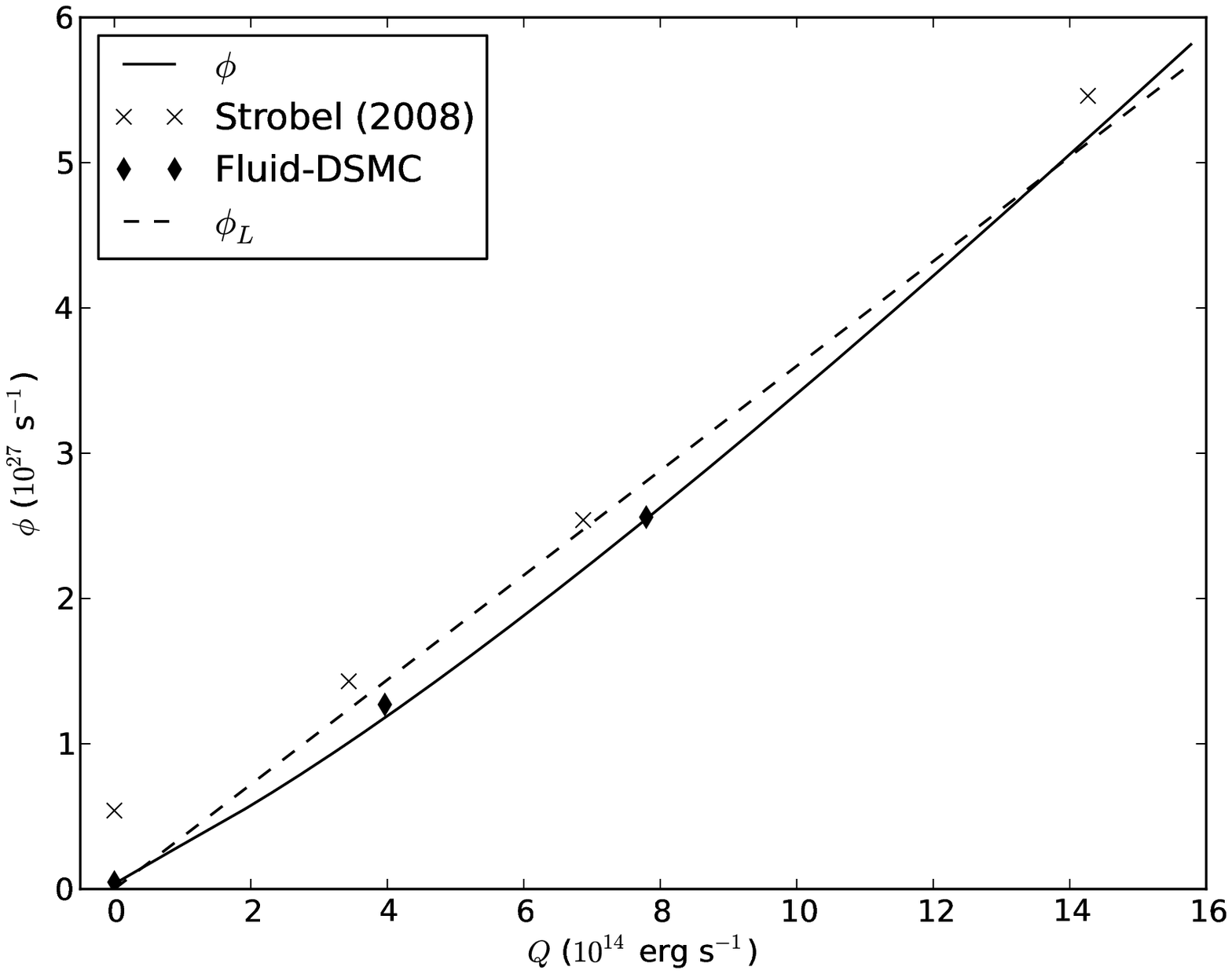}
  	\includegraphics[width = 0.49\textwidth]{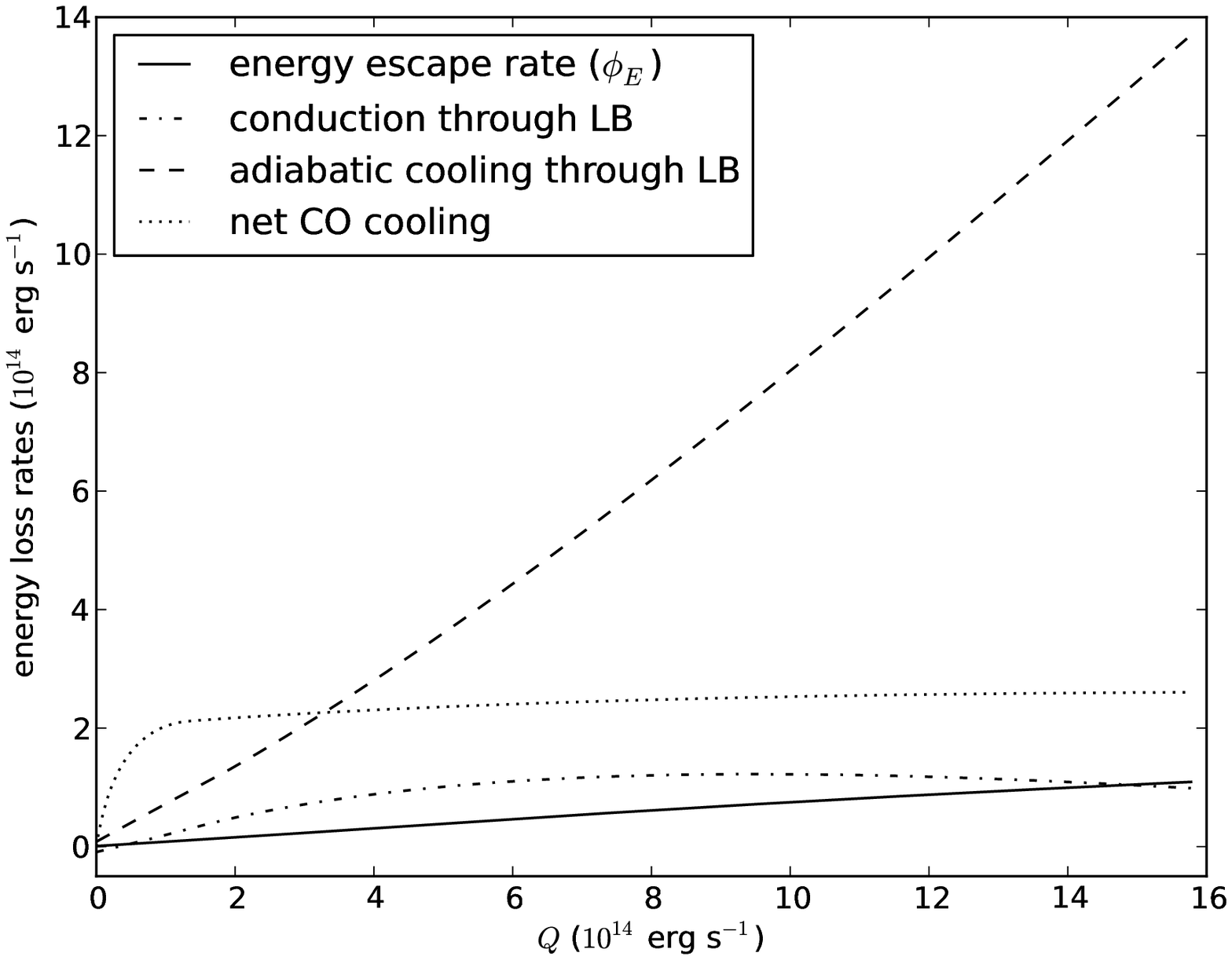}
 	\caption{(Left) Escape rate $\phi$ (solid line) plotted versus net heat deposited $Q$ from the fluid-Jeans model. Overlaid are escape rates from the fluid-DSMC model as black diamonds, and from the SHE model \citep{strobel2008} as x's. Also plotted is the energy limited escape rate $\phi_L=Q/U(r_0)$ (dashed line). (Right) Comparison energy loss mechanisms versus net heat deposited from the fluid-Jeans model.}
 	\label{fig:fjeans_escape}
\end{figure*}

In Figure~\ref{fig:fjeans_escape}, the escape rate versus the net heating rate for the fluid-Jeans model is compared to the fluid-DSMC and SHE models of \citet{strobel2008}. The fluid-DSMC results are nearly the same as the fluid-Jeans results, thus confirming that the fluid-Jeans model can be used to estimate the escape rate for this range of Jeans parameters. With the exception of the zero heating case, the SHE model results show a similar trend but are 43\%, 16\% and 7\% higher than the fluid-Jeans for solar minimum, mean, and maximum, respectively, with the latter approaching blow-off.

Also shown in Figure~\ref{fig:fjeans_escape} is the energy-limited escape rate compared to the fluid-Jeans escape rate. The good agreement between the two escape models suggests that most of the energy being deposited is going into removing the molecules from the gravitational well. The sources of cooling besides adiabatic expansion and loss to escape are conduction through the lower boundary and CO cooling (already included in $Q$) both of which are bounded, as found in the hydrodynamic model \citep{strobel2008}. The slope of the energy-limited escape rate can be better matched to our simulations by accounting for the internal energy (i.e. $C_p(T_0-T_x)$). The vertical offset can be partly compensated for by subtracting the loss to conduction through the lower boundary from the total energy deposited, but since it is bounded it is a small correction to the escape rate.

We also compared escape rates from the fluid-Jeans model, the hydrodynamic models of \citet{hunten1982} and \citet{mcnutt1989}, and the energy-limited escape model of \cite{lammer2009} for the same solar flux. For a consistent comparison we adopt a single solar heating source for methane with a globally averaged incoming solar energy flux, $F^\infty$, of 1$\times10^{-3}$ and 4$\times10^{-3}$ erg cm$^{-2}$ s$^{-1}$ (representing near solar minimum and maximum conditions respectively), and an absorption cross-section for CH$_4$ of 1.8$\times10^{-17}$ cm$^2$.  We assume a primarily nitrogen atmosphere with the values for conductivity used in this paper, $\kappa_0 = 9.38\times 10^{-5}$ J m$^{-1}$ K$^{-1}$, and a fixed $CH_4$ mixing ratio of 3\% when calculating the optical depth.

\begin{table}[h]
 \centering
 \begin{tabular}{lccc}
 	\hline\hline
			\multicolumn{4}{c}{$S=1\times 10^{-3}$ erg cm$^{-2}$ sec$^{-1}$} \\
 							& $\phi$ (10$^{27}$ s$^{-1}$)	& $r_h/r_0$	& $Q$ (10$^{14}$ erg s$^{-1}$) 	\\ \hline
	Fluid-Jeans		& 1.06		& 1.22		& 4.68	\\
	Watson			& 15.4		& 3.66		& 35.4	\\
	McNutt				& 0.84		& 1.17		& 3.69	\\
	ELE (i)			& 1.69		& -			& 4.68	\\
	ELE (ii)			& 1.66		& 1.32		& 4.62	\\
	ELE (iii)			& 1.41		& 1.22		& 3.94	\\ 
	ELE (iv)			& 0.95		& 1.0		& 2.64	\\ [0.5ex]
	\hline\hline
			\multicolumn{4}{c}{$S=4\times 10^{-3}$ erg cm$^{-2}$ sec$^{-1}$} \\
 							& $\phi$ (10$^{27}$ s$^{-1}$)	& $r_h/r_0$	& $Q$ (10$^{14}$ erg s$^{-1}$) 	\\ \hline
	Fluid-Jeans		& 8.46		& 1.25		& 23.3	\\
	Watson			& 32.1		& 2.73		& 76.8	\\
	McNutt				& 3.22		& 1.28		& 17.4	\\
	ELE (i)			& 8.40		& -			& 23.3	\\
	ELE (ii)			& 7.30		& 1.39		& 20.3	\\
	ELE (iii)			& 5.91		& 1.25		& 16.4	\\ 
	ELE (iv)			& 3.81		& 1.0		& 10.6\\ [0.5ex]
	\hline
 \end{tabular}
 \caption{Comparing the fluid-Jeans model, \citet{hunten1982}, \citet{mcnutt1989} and energy-limited escape (ELE). For each model, the values for each given are the escape rate $\phi$, the location of the heating layer in terms of the lower boundary $r_h/r_0$ (in the case of the fluid-Jeans model we give the location of the peak heating, which is where the slant optical depth is unity), and the net heating $Q$ are shown. The ELE model was applied in 4 ways: (i) using $Q$ from the fluid-Jeans model, (ii) $r_h$ is altitude of unit parallel optical depth, (iii) $r_h$ is the altitude of unit vertical optical depth, (iv) and $r_h$ is $r_0$ as if no estimate were given.}
 \label{tab:comp_esc}
\end{table}


In Table~\ref{tab:comp_esc}, neither of the simple hydrodynamic models produces a similar escape rate to the fluid-Jeans model for either heating case. The \citet{hunten1982} model is an upper bound on the escape rate, so it will consistently overestimate the escape rate.  The \citet{mcnutt1989} is a good estimate for the smaller heating rate, giving a similar estimate of both the escape rate and the altitude of the heating layer; however, it is quite different from the fluid-Jeans for the solar maximum case. While the hydrodynamic models can give good estimates of the escape rate for certain cases, the atmospheric structure of the solutions are quite different from that found the fluid-Jeans or fluid-DSMC models.

The energy-limited escape model can be a reasonable fit if the net heat deposited in the atmosphere is known. Though without detailed modeling, only an estimate of the amount of energy deposited via heating can be obtained. In Table~\ref{tab:comp_esc} we consider using the energy-limited escape model with 4 estimates of the net amount of heat deposited: i) the net heating rate is the net heating deposited in fluid-Jeans solution, ii) the net heating rate is $Q=\pi r_h^2 F^\infty$ with the radius of the heating layer ($r_h$) being the radius of unit parallel optical depth (i.e. slant column optical depth with $\theta=90^\circ$) found using the fluid-Jeans model of the atmosphere, iii) with the radius of the heating layer being the radius of unit vertical optical depth found using the fluid-Jeans model (this is also the altitude of the peak UV heating), iv) with the heat deposited at the lower boundary, $r_h=r_0$. For the first case, where $Q$ is known from the fluid-Jeans model, the energy-limited escape provides a good estimate for the highly heating case, but for the near solar minimum case it overestimate the escape as a significant amount of heat is conducted through the lower boundary. Using method (ii), the parallel optical depth provides a good estimate of the heat deposited in both case, but similar to method (i) it does not provide a good estimate of the escape rate for the solar minimum case as a significant amount of this heat is conducted through the lower boundary instead of driving escape. Methods (iii) and (iv) fail to predict $Q$ as accurately as using the radius of unit parallel optical depth.

We investigated the sensitivity of the fluid-Jeans model to changes in some of the model parameters. If the atmosphere were a monatomic gas ($C_p = 5/2k_b$) rather than diatomic ($C_p = 7/2k_b$), then $E_{esc}$ is about halved, primarily due to no internal energy being carried off by the escaping gas; however, the other properties are only significantly affected when the heating exceeds solar mean conditions. For solar mean, the escape rate increases by $\sim$1\% and the exobase is lower by $\sim$10 km. In hydrodynamic models it is often assumed that the energy carried off by escape is negligible and a positive non-zero energy loss can only decrease the escape rate \citep{watson1981}. Using $\phi_E=0$ (keeping $\phi$ as the Jeans value), then the escape rate increases by $\sim$5-10\% depending on solar flux. When the solar heating is at the solar max, corresponding to $\phi_E$ begin a comparable energy loss process to conduction or CO cooling, the exobase altitude is lowered from 14200 to 12600km, and the exobase temperature raised from 68 to 81K. Therefore, even though neglecting $\phi_E$ still produces a reasonable estimate of escape rate, it should be included when interested in the vertical structure of the atmosphere.

Another consideration is our definition of the Knudsen number used to determine the exobase altitude. The relevant length scale in atmosphere is determined by the density gradient $H_n = -n/(dn/dr)$. The isothermal approximation $H = k_bT/mg(r)$ was used in the above simulations. Using $\Kn=\ell/H_n$ to determine the exobase altitude results in a small correction (since $H\simeq H_n$ throughout the atmosphere) and has a negligible effect on the solution and escape rate. Since the fluid-Jeans atmosphere is highly extended one might consider $\Kn = \ell/r$, which is appropriate for comets, where the density drops off rapidly as $1/r^2$. Using this definition, the exobase altitude increases to 4.7, 9.3, 13.3, and 21.6 times $r_p$ for the four heating cases considered above. These values are about 50\% larger than those in Table~\ref{tab:fjeans} and \ref{tab:fjeans_max} and do not compare well with the fluid-DSMC results. For solar maximum such an estimate would place the exobase beyond the orbit of Charon.

\subsection{Similar Escape Rates for Small Jeans Parameters}

This study has shown that the escape rates of the fluid-Jeans and fluid-DSMC solutions are close to those found in the hydrodynamic models when there is heating, and become closer as the heating increases.  We posit that this is due to the Jeans parameter being lowered at the upper boundary of the heating layer as the heating rate increases, and that for small Jeans parameters only a small range of escape rates lead to valid solutions as discussed below.

\begin{figure*}[t]
   \centering
   \includegraphics[width=.7\textwidth]{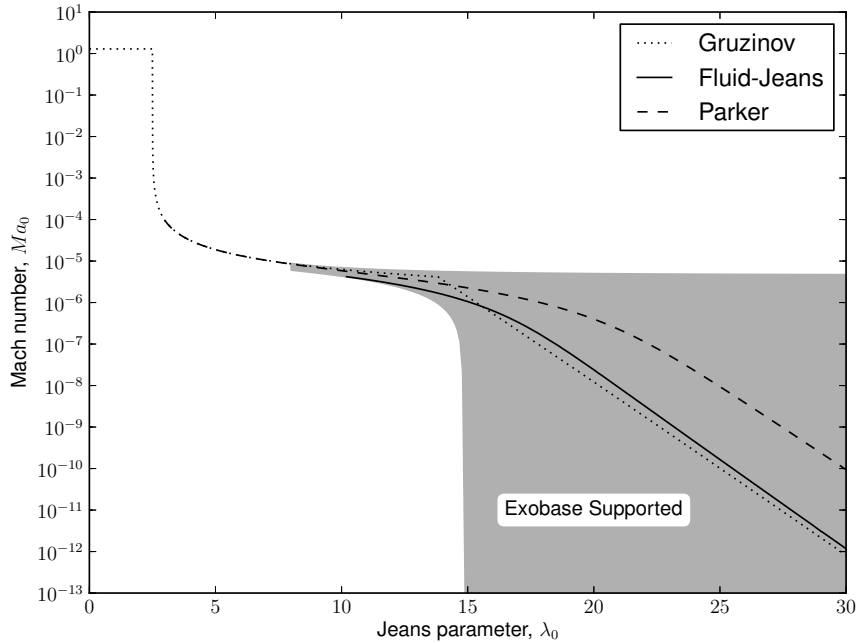} 
   \caption{Escape rate solutions from \citet{gruzinov2011} and \citet{parker1964a}, compared to our fluid-Jeans solutions as a function of Jeans parameter without heating for $\Kn_0=10^{-5}$ (about Pluto's 1450 km value). The shaded region is the range of escape rates that support an exobase.}
   \label{fig:Ma_versus_lambda}
\end{figure*}

Figure~\ref{fig:Ma_versus_lambda} shows the Mach number at the lower boundary (which is proportional to the escape rate) as a function of the Jeans parameter at the lower boundary for the \citet{parker1964a} hydrodynamic model, \citet{gruzinov2011} analytic model, and fluid-Jeans model of escape when all the heat is deposited below the lower boundary. The analytic model of \citet{gruzinov2011} resulted in an escape rate that compared well with the DSMC simulations of \citet{volkov2011} through the transition from hydrodynamic blowoff (small Jeans parameters) to Jeans escape (large Jeans parameters). 

It is seen in Figure~\ref{fig:Ma_versus_lambda} that the Parker and fluid-Jeans solutions begin to converge as the Jeans parameter at the lower boundary decreases. As the heating increases, the atmosphere expands and the Jeans parameter above the heating layer decreases, so the two models should produce similar escape rates. The fluid-DSMC solution lies between the Parker and fluid-Jeans solutions as it is enhanced relative to Jeans but not a sonic solution.

Solutions to the full hydrodynamic equations of \citet{parker1964a} can be very sensitive to the escape rate, especially for low Jeans parameters, and not all choices of escape rate lead to valid solutions. The shaded region in Figure~\ref{fig:Ma_versus_lambda} gives the range of escape rates for which the atmosphere eventually attains an exobase. This region was determined by computing the solutions to the fluid equations for choices of $\Ma_0$ and $\lambda_0$ to see if an an exobase was reached. Note that this region significantly narrows for $\lambda_0 \lesssim 15$ for this value of $\Kn_0$. 




Both the fluid-Jeans and Parker's solution lie within this region, as does Gruzinov's solution and the SHE solution, which is close to Parker's. Therefore, for both solutions to have an exobase for small Jeans parameters they must have similar escape rates. Even though the escape rates are within an order of magnitude, the upper atmosphere can have very different structures, from a supersonic, rapidly cooling solution, to the near isothermal solution near the lower edge of the narrow region.

When the atmosphere is heated this argument must be applied above the heating layer. Performing similar analyses as that presented in Figure~\ref{fig:Ma_versus_lambda} we find the onset of the narrow region decreases in $\lambda_0$ with an increase in $\Kn_0$. Applying this at $\Kn_0=0.1$, where the heating has been cut-off, the narrow region occurs for $\lambda_0 \lesssim 6.7$. We find that this value is achieved near solar mean heating.  Therefore for solar mean and maximum heating the SHE model and our fluid-DSMC and fluid-Jeans models produce similar escape rates, even though the solutions are very different in describing the extended atmosphere.

\section{Conclusion}

We compared two methods (fluid-DSMC and fluid-Jeans) for modeling the upper thermosphere and characterizing atmospheric escape. Having shown earlier that such models can be scaled \citep{volkov2011}, we considered Pluto's atmosphere as an example because of the anticipated NH encounter. We found that, in the presence of significant heating, the fluid-Jeans total escape rate did not change significantly from that obtained from recent hydrodynamic models. However, many other characteristics of the atmosphere are affected, namely the exobase altitude and the existence of a sonic point.

The fluid-DSMC model requires no assumption on upper boundary conditions, we used this as a baseline for testing fluid models that require assumptions about the particle and energy flux at the upper boundary. By iteration, a consistent fluid-DSMC describes the escape rate and energy escape rate as well as the density, temperature and vertical flow speed out to 10's of planetary radii. The escape rate found is approximately 2 times Jeans escape rate, but this value changes slowly with the amount of heat deposited in the upper atmosphere, as well as with the temperature at the lower boundary of the simulations. The arrival of NH should occur when the FUV/EUV flux is between the solar minimum and mean cases considered in this paper. 

The fluid-Jeans model uses the Jeans escape for an upper boundary condition as an approximation to the DSMC model in the exobase region. The resulting escape rate of the fluid-Jeans model is surprisingly close to that given by the fluid-DSMC model; the small differences in escape rates do not imply that the temperature and density structures are similar in the two models. By scaling the Jeans escape rate by a factor between 2 and 3 the results become closer to the fluid-DSMC.

We also showed that the energy-limited escape rate, often used for estimating atmospheric loss from exoplanets, was surprisingly accurate over a broad range of heating conditions from solar minimum to solar maximum. Since the structures of exoplanet atmospheres are usually not known, we also show that in applying this estimate one should use a radius associated with the FUV/EUV absorption radius or, better, use that radius for which the horizontal line of sight optical depth is unity. This radius can be roughly estimated by scaling the visible radius using the ratio of the cross sections, the visible light scattering cross section to the FUV/EUV absorption cross, along with a scale height estimated using the equilibrium temperature. In this way, the results can be applied when modeling the evolution of exoplanet atmospheres.

For the heating rates considered here we have not seen the transition from Jeans to hydrodynamic escape \citep[e.g.,][]{johnson2013}; however, the simulated highly extended atmosphere for NH encounter conditions means that the direction of the modeling effort for Pluto needs to change. That is, thermal escape calculations generally terminate the energy deposition at very high altitudes as the heating efficiency of the background gas due to solar radiation absorption drops rapidly with increasing altitude in the exobase region. However, the molecules in the upper region of a highly extended atmosphere \textit{do absorb} radiation. Although this might not efficiently heat the atmosphere, a fraction of it does produce non-thermal escape, as described for other bodies \citep{johnson2008}. In addition, the interaction with the solar wind occurs at much larger distances from Pluto than previously predicted and can more efficiently strip the atmosphere \citep{bagenal1989}. For the extended atmosphere described here, these non-thermal processes at the NH encounter conditions could be comparable to thermal escape estimates and strongly affect the NH observations. Such modeling is now in progress.

\section*{Ackowledgements}
This research was supported by the NASA Planetary Atmosphere Program, the NSF Astronomy Program, and the Virginia Space Grant Consortium. We acknowledge helpful comments and advice from colleagues L. Young, A Volkov, and J. Deighan.

\appendix
\section{An implicit time stepping scheme with varying step size}
In this appendix, we describe the numerical equations and techniques used to solve the diffusion equation on an unequally spaced grid. We define a grid $r_m$, where $m=1, \ldots, M$ and a spacing $\Delta r_m = r_{m+1}-r_m$. The average grid spacing  at $r_m$ as $\overline{\Delta r}_m = (r_{m+1}-r_{m-1})/2$. Finally, the finite difference approximation to $\xi(r_m)$ as $\xi_m$.

First we describe approximate the derivatives $\xi^\prime(r)$ and $\xi^{\prime\prime}(r)$ using the unequally spaced grid. Using the points $\xi_{m+1}$, $\xi_m$, $\xi_{m-1}$ to approximate $\xi^\prime(r)$, one might use a central difference scheme $\xi^\prime(r) \approx (\xi_{m+1}-\xi_{m-1})/(2\overline{\Delta r})$; however, this scheme is not second order accurate, $O(\overline{\Delta r}^2)$, as it would be in the case of equal spacing. The correct error term is proportional to $(\Delta r_m-\Delta r_{m-1})\xi^{\prime\prime}(\tilde{r})$. While the quantity $(\Delta r_m-\Delta r_{m-1})$ might be small compared to $(\overline{\Delta r}_m^2)$, the effect of this error in the model is to overestimate the conductivity. To correct for this we use the finite differences:
\begin{align}
	\xi^\prime(r)&= 
		\frac{\Delta r_{m-1}^2(\xi_{m+1}-\xi_m) + \Delta r_m^2(\xi_m-\xi_{m-1})}
		{2 \Delta r_m \Delta r_{m-1} \overline{\Delta r_m}}
		\label{eq:dxidr}\\
	\xi^{\prime\prime}(r) &=
		\frac{\Delta r_{m-1}(\xi_{m+1}-\xi_m) - \Delta r_m(\xi_m-\xi_{m-1})}
		{\Delta r_m \Delta r_{m-1} \overline{\Delta r_m}}
		\label{eq:d2xidr2}
\end{align}
The error term on $\xi^\prime$ is $\Delta r_m\Delta r_{m-1} \xi^{\prime\prime\prime}(\tilde{r})$, which is second order accurate and does not affect the conductivity. The error term for $\xi^{\prime\prime}$ is $(\Delta r_m-\Delta r_{m-1})\xi^{\prime\prime\prime}(\tilde{r})$. While the finite difference approximation for $\xi^{\prime\prime}$ is not second order accurate, this is the minimal error term one can get from the 3 points used. It is important that we capture conduction correctly as to ensure a constant energy flux through the upper domain where there is zero heating and the grid spacing is large.

We adapt the implicit scheme proposed by \citet{zalucha2011}, using the same notation, to solve the diffusion equation
\begin{equation}
	\frac{\partial \xi}{\partial t} = S + B \xi + C \frac{\partial \xi}{\partial r} + G \frac{\partial^2\xi}{\partial r^2} 
	\label{eq:diffeq}
\end{equation} 
on a domain $r \in (r_0,\ r_{t})$. The coefficients $B$, $C$, $G$ and source term $S$ are functions of $r$ and $t$.  The boundary conditions for the fluid-Jeans model can be expressed as
\begin{align}
	\xi(r_0) &= \xi_0\\
	\left. \frac{d\xi}{dr} \right|_{t} &= F_{t} \label{eq:dxidrtop}
\end{align}
We make two modifications to Zalucha's scheme. First, the finite difference expressions for an unequally spaced grid defined in Eqs. (\ref{eq:dxidr}) and (\ref{eq:d2xidr2}) are substituted into (\ref{eq:diffeq}), and second, the finite difference formulation at the upper boundary will satisfy the energy flow restriction of Eq. (\ref{eq:dxidrtop}) as well as the diffusion equation (\ref{eq:diffeq}).

Implementing the time grid $t_i = i\Delta t$ for $i = 0, 1, \ldots$ and the notation $\xi(r_m,t_i) = \xi_{m}^{i}$. An implicit scheme is obtained by substituting the finite differences, Eqs. (\ref{eq:dxidr}) and (\ref{eq:d2xidr2}), evaluated at the advanced time step into the diffusion equation (\ref{eq:diffeq}). For the time derivative, we use $(\xi_m^{i+1}-\xi_m^i)/\Delta t$. Solving for the advanced time step then requires solving a tridiagonal system of equations
\begin{equation}
	\alpha_m\xi_{m-1}^{i+1} + \lambda_m\xi_{m}^{i+1} + \omega_m\xi_{m+1}^{i+1} = q_m
\end{equation}
The lower boundary condition $\xi_{1}^{i} = \xi_0$ is then expressed as $\alpha_1=0$, $\lambda_1=1$, $\omega_1=0$, $q_1=\xi_0$.  The interior point (m = 2,\ldots,M-1) are expressed as
\begin{equation}
	\left. \begin{aligned}
	 \alpha_m &= \frac{\Delta t}{\Delta r_{m-1}\overline{\Delta r}_m}
		\left(\frac{\Delta r_m}{2}C_m-G_m\right) \\
	\lambda_m &= 1-\Delta t \left(B_m + \frac{1}{\Delta r_m\Delta r_{m-1}} \right.\cdot \\
		&\hspace{0.5cm}\left.\vphantom{\frac{1}{\Delta r_m\Delta r_{m-1}} }\left( (\Delta r_m-\Delta r_{m-1}) C_m - 2G_m \right) \right) \\
	\omega_m &= \frac{-\Delta t}{\Delta r_m\overline{\Delta r}_m}
		\left(\frac{\Delta r_{m-1}}{2}C_m+G_m\right) \\
	 q_m &= \xi_{m}^{i} + \Delta t S_m
	\end{aligned}\right\}
\end{equation}

The procedure for interior points is modified for to enforce the condition on $\frac{d\xi}{dr}$ at the upper boundary. Eq.~(\ref{eq:dxidrtop}) is substituted directly into Eq.~(\ref{eq:diffeq}) for $\xi^\prime(r)$. Using Eq.~(\ref{eq:d2xidr2}) with $m=M$ for $\xi^{\prime\prime}(r)$ at the upper boundary introduces the phantom point $\xi_{M+1}$, but by equating Eq.~(\ref{eq:dxidr}) to Eq.~(\ref{eq:dxidrtop}) we can remove it from the equation. The result is expressed as 
\begin{equation}
	\left. \begin{aligned}
	\alpha_M &= -\Delta t \frac{1+\Delta r_M/\Delta r_{M-1}}{\Delta r_{M-1}\overline{\Delta r}_M} G_M\\
	\lambda_M &= 1-\Delta t \left(B_M - \frac{1+\Delta r_M/\Delta r_{M-1}}{\Delta r_{M-1}\overline{\Delta r}_M} G_M \right)\\
	\omega_M &= 0 \\
	q_M &= \xi_{M}^{i} + \Delta t \left( S_M +F_{top}\left( C_M+\frac{2}{\Delta r_{M-1}}G_M \right)\right)
	\end{aligned}\right\}
\end{equation}

\singlespace
\bibliography{pluto_bib}

\end{document}